\documentclass[12pt]{iopart}
\usepackage{amssymb,graphicx,amsbsy}
\begin{document}

\title{The Ten Thousand Kims}
\author{Seung Ki Baek$^1$, Petter Minnhagen$^1$, Beom Jun Kim$^{2,3}$}
\address{$^1$Integrated Science Laboratory, Department of Physics, Ume{\aa} University, 901 87 Ume{\aa}, Sweden}
\address{$^2$BK21 Physics Research Division and Department of Physics, Sungkyunkwan University, Suwon 440-746, Korea}
\address{$^3$Asia Pacific Center for Theoretical Physics, Pohang 790-784, Korea}
\ead{Petter.Minnhagen@physics.umu.se}

\begin{abstract}
In the Korean culture the family members are recorded in special family
books. This makes it possible to follow the distribution of Korean family
names far back in history. It is here shown that these name distributions
are well described by a simple null model, the random group formation (RGF)
model. This model makes it possible to predict how the name distributions
change and these predictions are shown to be borne out. In particular, the
RGF model predicts that, for married women entering a collection of family
books in a certain year, the occurrence of the most common family name
``Kim'' should be directly proportional the total number of married women
with the same proportionality constant for all the years. This prediction is
also borne out to high degree. We speculate that it reflects some inherent
social stability in the Korean culture. In addition, we obtain an estimate of
the total population of the Korean culture down to year 500 AD, based on the
RGF model and find about ten thousand Kims.
\end{abstract}


\maketitle

\section{Introduction}

Your family name is very important in the Korean culture. Abandoning your
family name is extremely unusual and considered dishonorable. A common
metaphor is to pledge your family name to a given promise. The importance
is also reflected in the fact that a married woman carries the family name
of the family she comes from. In addition, the Confucian tradition has
encouraged a family to record the genealogical tree in special books, which
then records the women's family names flowing into the family book by
marriages. The children inherit the name of the father. Some of the family
books go back more than 500 years.
 
The distribution of family names is often described in terms of the
probability $P(k)$ to randomly pick a family name which occurs $k$ times
within the population. The frequency distribution $P(k)$ for family names
are examples of broad `fat-tailed' distributions which, at least crudely,
can be described by  power laws $P(k)\propto 1/k^\gamma$, as has been
studied before~\cite{miya,zane,reed}. In particular, Korean family names
have a very broad distribution with a $\gamma$ close to one~\cite{fam0}. It
is common to try to connect the approximate power-law form of family
distributions to growth models of the total population
\cite{miya,zane,reed,fam1}. Such models usually yield power laws with
approximately the Zipf´s law exponent $\gamma=2$, i.e, $P(k)\propto 1/k^2$.
However, this does not describe the Korean family distribution, which has a
much slower falling-off at large $k$. It has been suggested that this is
because the rate of introducing new family names in Korea is very
slow~\cite{reed,fam1}. In accordance with this, it was in \cite{fam1}
shown that that a growth model with an introduction of family names which
approaches zero can indeed yield a power law with $\gamma=1$ instead of
$\gamma=2$.

In \cite{zipf}, it was shown that system-specific growth models are
usually too restrictive to catch some of the essential characteristic
features of frequency distributions. Examples are the dependence of $\gamma$
on the size of the data set and the connection between the data-set size and
the size of the largest frequency. In the present paper, we reinvestigate
the historical Korean family books from this particular perspective. We use
a collection of Korean family books to estimate the change in the frequency
distribution of family names for the last five hundred years. We then
compare these changes with the predictions of the Random Group Formation
(RGF) model introduced in \cite{zipf}. The RGF model assumes maximal
mixing for each given data size and it is shown that the predictions from
this model are borne out to a striking degree. In particular, it is found
that the proportion of persons named `Kim' is constant irrespective of all
social changes, wars, earthquakes, famines, plagues, fertility variations,
industrial revolution, etc. and this constancy is also an inherent feature of
the RGF model.
 
In \sref{sec:books}, we describe how we use the data from the Korean
Family books and in \sref{sec:rgf}, we give a brief recapitulation of
the Random Group Formation model. The comparison between data and
theoretical predictions are given in \sref{sec:analysis}, while
\sref{sec:conclude} contain some concluding remarks.

\section{Korean Family Books}
\label{sec:books}
Our data is extracted from the ten Korean family books which were also
analyzed in \cite{fam2}. The data we extract in the present
investigation is the total number $M$ of married women, who were registered
with marriage year into these ten books during a specific 30-year period
between 1510 and 1990. For each period, the number of different family names
$N$ and the number of women having the most common family name (usually
`Kim'), $k_{\rm max}$, are also extracted. The results are given in
\tref{table:book} which contains sixteen historical windows. This
data set is analyzed and compared with census data for the whole Korean
population from year 2000 (see \cite{fam0}).

\begin{table}
\caption{Statistical quantities of the family books analyzed in this work.
Here $M$ means the total number of women entering this family by marriage at
the specified period. $N$ means the number of different family names that
the women carried. Among these $N$ different names, we find the one with
the largest number of carriers and denote this number by $k_{\rm max}$.}
\begin{tabular*}{\hsize}{@{\extracolsep{\fill}}ccccc}\hline\hline
$\sharp$ & period & $M$ & $N$ & $k_{\rm max}$\\\hline
1  & 1510 -- 1540 &  33     &   19 &    6      \\
2  & 1540 -- 1570 &  94     &   31 &    23     \\
3  & 1570 -- 1600 &  215    &   38 &    40     \\
4  & 1600 -- 1630 &  384    &   48 &    88     \\
5  & 1630 -- 1660 &  643    &   52 &    119    \\
6  & 1660 -- 1690 &  1039   &   59 &    230    \\
7  & 1690 -- 1720 &  1534   &   74 &    311    \\
8  & 1720 -- 1750 &  2313   &   82 &    407    \\
9  & 1750 -- 1780 &  3524   &   83 &    598    \\
10 & 1780 -- 1810 &  5640   &   96 &    1069   \\
11 & 1810 -- 1840 &  8499   &  110 &    1502   \\
12 & 1840 -- 1870 &  13028  &  114 &    2256   \\
13 & 1870 -- 1900 &  19559  &  129 &    3410   \\
14 & 1900 -- 1930 &  37531  &  153 &    6888   \\
15 & 1930 -- 1960 &  79935  &  163 &    15735  \\
16 & 1960 -- 1990 &  47554  &  162 &    9693   \\\hline\hline
\end{tabular*}
\label{table:book}
\end{table}

As in \cite{fam2}, we argue that the statistics of this collection of
women's family names should bear a strong resemblance with the whole
population during the same period in the following sense: suppose that out
of the whole population of women who got married at a certain period, you
have randomly selected $M$ women. There is a certain chance that a chosen woman
has a family name which occurs $k$ times among the $M$ picked women. Suppose
that the probability distribution for the  frequency of different family
names within a group of $M$ randomly selected women is $P_M(k)$. Then the
$M$ selected women on the average have family names which occur $M/N$ times
since $\sum_{k=1}^{k_{\rm max}}kP_M(k) = \left< k \right>=M/N$, where
$k_{\rm max}$ is the most frequent
name. The data for an entry in \tref{table:book} corresponds to a single
try of choosing $M$ women out of the whole collection of women who got
married during the period. Suppose that you have now instead picked $M$ random
persons out of the whole population. Then provided that the married women were
really randomly distributed over the population, the result would be
statistically the same: the probability distribution for the frequency of
different family names within a group of $M$ randomly selected persons would
again be given by $P_M(k)$.

There are two important points to notice in this context. The first is that
$P_M(k)$ for $M$ randomly selected persons does not have the same functional
form as $P_{M_{\rm tot}}(k)$ for the complete population $M_{\rm
tot}$~\cite{zipf}.
In other words, the family-name distribution depends on the size of $M$. The
second is that $M_{\rm tot}(t)$ depends on time $t$ in an unknown (but
presumably rather nontrivial) way reflecting the history of the Korean
people. There is, at least \emph{a priori}, no obvious relation between the
inflow of married women into the ten specific family-name books on one
hand, and the total population of the Korean people on the other: A
prosperous family could have a large inflow of married women even in periods
when the total population decreases.
 
\section{The RGF model}
\label{sec:rgf}
The random group formation model tries to catch the essential features of
the group-size distribution when $M$ objects are divided into $N$ groups by
assuming optimal mixing~\cite{zipf}. In the case of family names, the persons
are the objects and the groups are formed by the persons carrying the same
family name. The RGF model does not make any explicit assumption about what
particular process is responsible for the creation of the groups. Instead it
is assumed that, whatever this process might be, the result is that the
optimal mixing condition is on the average  approximately fulfilled at all
times. The optimal mixing condition corresponds to a maximum entropy
condition for the group-size distribution $P_M(k)$. The appropriate maximum
condition can be formulated in terms of a maximum mutual-information
principle or equivalently as a minimum information-cost
condition~\cite{cover,zipf}. The result is a distribution function of the
explicit form
\begin{equation}
P_M(k)= A\frac{\exp(-bk)}{k^\gamma}.
\label{eq:pk}
\end{equation}
where $P_M(k)$ obeys the following set of self-consistent equations:
\[ \left \{ \begin{array}{l}
\sum_{k=k_0}^{M} A\frac{e^{-bk}}{k^\gamma}=1\\
\sum_{k=k_0}^{M} A k\frac{e^{-bk}}{k^\gamma}=M/N
\label{eq:basic}
\end{array} \right.\]
where $k_0$ is the size of the smallest group. In the present
investigation, this limit is always $k_0=1$, but in other applications it
can be generalized to an arbitrary $k_0$. This means that the constants
$\gamma$ and $b$ are interdependent through the relation:
\begin{equation}
\frac{\sum_{k=k_0}^{M}\frac{e^{-bk}}{k^{\gamma-1}}}
{\sum_{k=k_0}^{M}\frac{e^{-bk}}{k^\gamma}} = \frac{M}{N}.
\label{eq:b}
\end{equation}
The self-consistency condition connects the entropy of $P_M(k)$ to the size
of the largest group $k_{\rm max}$, through the relation
\begin{equation}
\left< k_{\rm max} \right> = \frac{\sum_{k=k_c}^{M} k P(k)}{\sum_{k=k_c}^{M}
P(k)}.
\label{eq:eps}
\end{equation}
where $\left< k_{\rm max} \right>$ is the average size of the largest group
and the value of $k_c$ is determined such that there is on the average only
a single group in the interval $[k_c,M]$, i.e., $\sum_{k=k_c}^{\infty} P(k)
= 1/N$, which means that \eref{eq:eps} gives the average $\left<k_{\rm
max} \right>$. When analyzing data we will approximate $k_{\rm max}$ for the
data with the average size of the largest group $\left<k_{\rm max} \right>$
obtained for the RGF model. This set of self-consistent equations yields a
unique solution $P_M(k)$ of the RGF form for a given values of $M$, $N$,
$k_{\rm max}$, and $k_0$~\cite{zipf}.
Suppose you have a collection of $M$ persons each carrying one out of $N$
different family names which are distributed according to the frequency
distribution $P_M(k)$. What is the corresponding frequency distribution
$P_m(k)$ for $m$ persons randomly picked from the original total $M$? In the
case that the persons are picked randomly, $P_m(k)$ is given by the
transformation
\[
P_m(k)=\frac{\sum_{k'=k} \left(\frac{m}{M-m} \right)^{k}
\left(\frac{M-m}{m}\right)^{k'} {k' \choose k}
P_M(k')}{1-\sum_{k'=1}(\frac{M-m}{M})^{k'}P_M(k')} 
\]
where ${k' \choose k}$ is the
binomial coefficient. In the context of word-frequency of books, this
transformation is sometimes referred to as the Random Book Transformation
(RBT)~\cite{tr,meta}. The crucial point is that, if you start with a certain
$P_M(k)$ of the RGF form of \eref{eq:pk}, then all of $\gamma$, $b$,
and $A$ will change with $m$: a random reduction of a data set is not
scale-invariant with respect to the frequency distribution and as a
consequence, the power-law index $\gamma$ changes~\cite{tr,meta,zipf}. Another
characteristic feature of the transformation is that the number of persons
in the largest family group is proportional to the size of the
data set~\cite{zipf}.

In order to predict the change of the RGF function under the data-size
reduction, we first numerically calculate $\left< k \right>_m =
\sum_{k=1}^{m}kP_m(k)$ using the RBT transformation and then use the RGF
self-consistent equations for the input values $m$, $n=m/\left<k\right>_m$,
and $k_{\rm max}(M)m/M$, to obtain the corresponding $P_m(k)$ of the form of
\eref{eq:pk}. In this way, one obtains a predictions for $P_m(k)$
starting from a known $P_M(k)$ when $m<M$. It is also possible to get
predictions for an increase of the data set (i.e. $m>M$) in a similar way.
However, in the present paper, we will, when predicting the increase of a
data set, resort to two approximate relations found in \cite{zipf}
\begin{equation}
\gamma(M) - 1 \propto \frac{1}{\log M},
\label{eq:ext}
\end{equation}
describing the expected approach to the large-$M$ limit together with the
approximate relation $b\propto 1/M$.

Earlier attempts to explain family-name distributions have usually been
connected to explicit assumption about the time evolution linked to the
growth of the population~\cite{miya,zane,reed,fam1}. One may then ask how
the time evolution enters the RGF model. The answer is that it only
enters indirectly; the RGF model is in itself history-independent and at
each time only depends on the instantaneous input parameters. However, one
of the input parameters is the size of the data set $M$. Suppose that $M$ is
the total population, then this parameter is indeed history- and
time-dependent. One might even suspect that it has a complicated
time dependence $M(t)$ reflecting changes due to wars, earthquakes, famines,
plagues, fertility variations, industrial revolution etc. The point is that
the RGF model assumes that, whatever this actual historical time dependence
might be, the resulting family-name frequency distribution to good
approximation on the average is given by the maximal mixing condition which
only depends on the instantaneous value of $M(t)$.

A parallel example of this is provided by the word-frequency distribution of
novels written by an author~\cite{meta,zipf}: no matter what size the novel
has or when it was written, to good approximation the word-frequency
distribution for a novel of an author only depends on the number of words it
contains~\cite{meta}. This size dependence is to very good approximation
given by the RGF model~\cite{zipf}. The fact that the word frequency of an
author to good approximation only depends on the size of the text is
equivalent to characterizing an author by a single very large ``meta-book''
from which the average frequency distribution from any text size written by
the author can be obtained~\cite{meta}.

\section{Analysis}
\label{sec:analysis}

\begin{figure}
\begin{center}
\includegraphics[width=0.45\textwidth]{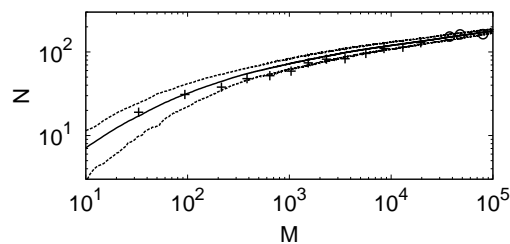}
\end{center}
\caption{Comparison between the expected number of different family names
and data. The full drawn curve is the number of different family names $N$
as a function of women's names $M$, when $M$ women are randomly chosen from
all the women's recorded in the ten books given in \tref{table:book}
during the period 1900-1990. The curve is the average over $10^2$ random
choices, and the dotted lines show three standard deviations.
The three open circles are the explicit data from \tref{table:book} for
the three periods 1900-1930,1930-1960, and 1960-1990.
The crosses represent the remaining historical data which decreases
monotonously with time. The data is consistent with a random drawing of
persons from a \emph{time-independent} distribution.}
\label{fig:nm}
\end{figure}

\begin{figure}
\begin{center}
\includegraphics[width=0.45\textwidth]{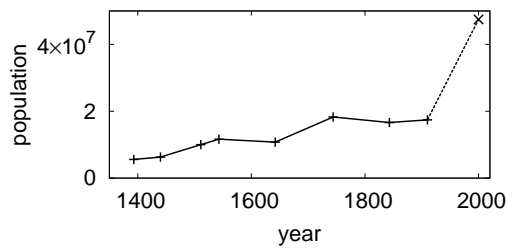}
\end{center}
\caption{Historical estimates of the Korean population in
\cite{history} (crosses). The rightmost point indicates the census data
in year 2000.}
\label{fig:estimate}
\end{figure}

In order to test the RGF model, we start with the three most recent entries
in \tref{table:book} which spans the time period 1900-1990. These three
entries  together contain $M_{1900-90}=165020$ women each having one family
name out of $N_{1900-90}=194$. Out of this data set, we randomly select
$M<M_{1900-90}$ women and calculate the average $N$ for each $M$. The full
curve in \fref{fig:nm} shows the average $N(M)$. This random selection
prediction based on the specific period 1900-1990 is compared to the data
covering all the time periods from 1500-1990. If the women flowing into the
entries by marriage are statistically equivalent to selecting random persons
in the total population and if the population was static in time, then the
agreement between the data and the random selection would be easy to
understand. However, the latter assumption is of course not true: both the
total population $M_{\rm tot}(t)$ and the number of different names
$N_{\rm tot}(t)$ change in response to historical developments : From about 1500
to 2000, the population in Korea increased by roughly a factor of 6, from
about 8 mil to 46 mil. This increase is linear in time up to 1900 and then
increases faster (compare \fref{fig:estimate}). During the same period, the
number of family names only increased very slowly, again with a sharp
increase around 1900. In the rough estimate given below, the increase is 
about 27 names from around 1500 to 2000. In spite of
this, the results shown in \fref{fig:nm} suggest that this time evolution
is such that $N(t)$ depends on time $t$ only through $M(t)$, such that at
all times $N(M)$ is a unique function. The uniqueness of the function $N(M)$
is also a consequence of the RGF model. In the context of word frequencies,
it corresponds to the meta-book concept discussed in \cite{meta,zipf}:
the word frequency used by an author is to good approximation given by a
unique function $N(M)$, where $N$ is the number of distinct words and $M$ is
the total number of words, independent of which book, what length the book
has, or when it was written. In short, \fref{fig:nm} suggests that $N$
is only a function of $M(t)$ and this particular feature is also consistent
with the RGF model.

\begin{figure}
\begin{center}
\includegraphics[width=0.45\textwidth]{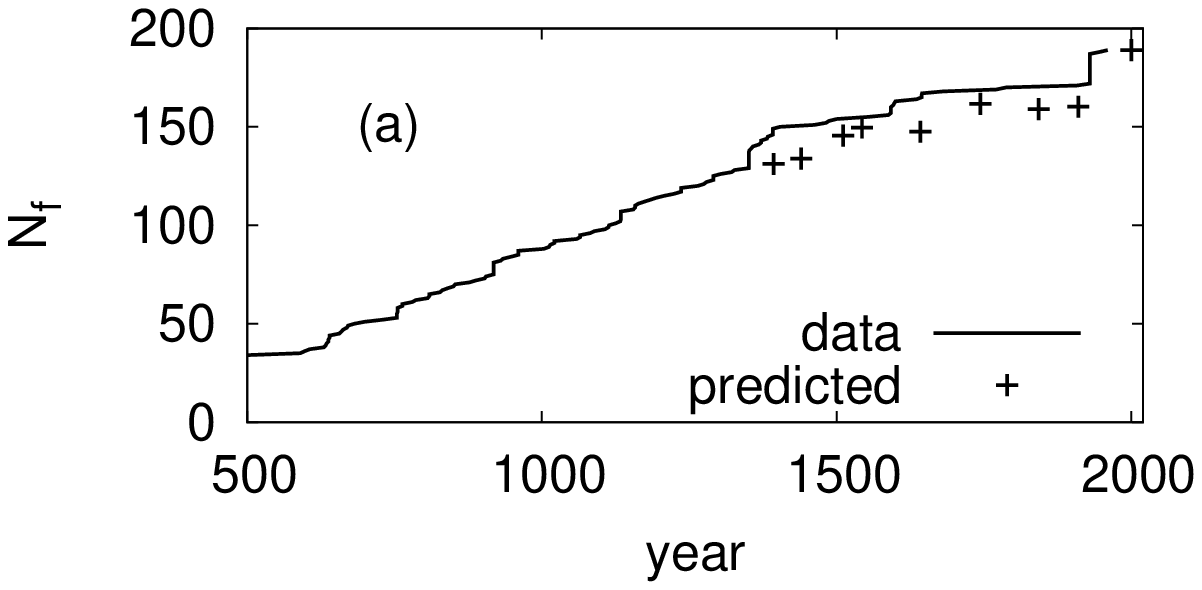}
\includegraphics[width=0.45\textwidth]{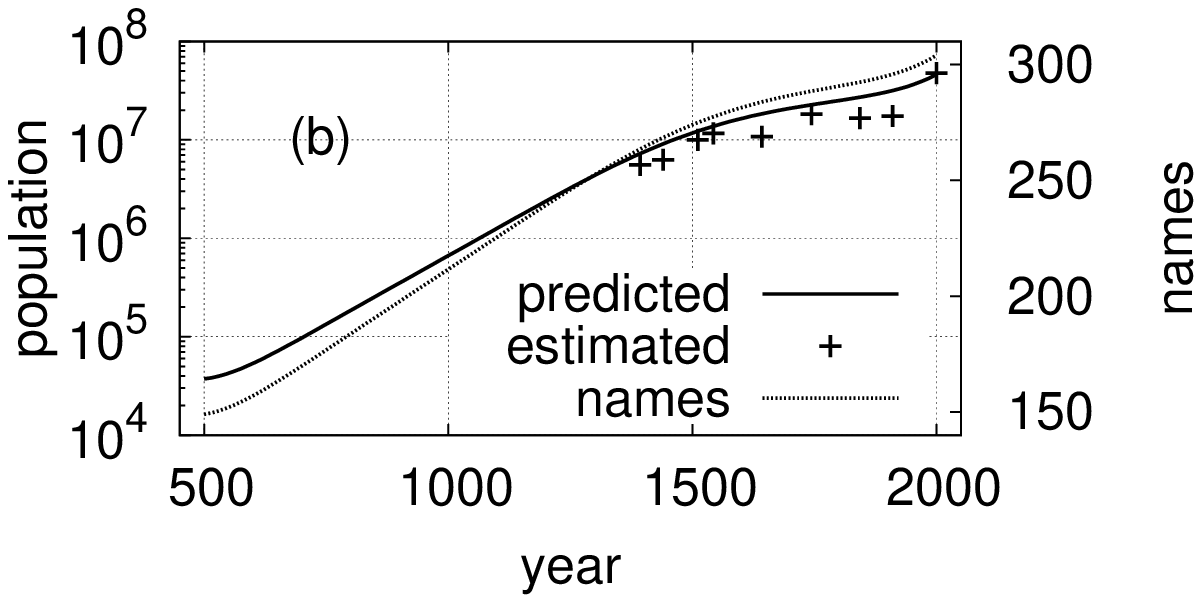}
\end{center}
\caption{(a) Number of family names that are known to be
introduced in Korea until a certain year (solid) and our predicted number of
them (crosses), which is obtained by using the historical estimates
[\fref{fig:estimate}] and the family books in 1900-1990 (see text).
(b) Our predicted size of population in the past (solid),
based on the number of people in the same family books, carrying
$N_f$ different names shown in (a). The crosses are the same historical
estimates as in \fref{fig:estimate}. Using this prediction as the input
parameter of the RGF description [see \eref{eq:ext}], we also estimate the
\emph{total} number of family names (dotted).
}
\label{fig:history}
\end{figure}

In order to illustrate the consequences of a time-independent distribution
further, we use the data from \cite{fam1} for the introduction years of 189
families. \Fref{fig:history}(a) shows the number of these 189
families which existed at a given time from year 500 to 2000. Note that the
increase is slow: from year 1500 to 2000 the increase is only about 20\%.
Now imagine that you pick a group of male persons which belong to one
of these 189 families in year 2000. If we follow the lineage of this group of
$m$ persons back in time, the only decrease in the number of family names is
caused by the fact that the lineage stops. Suppose that you randomly pick a
fraction $m/M$ of the population for year 2000, which on the average
contains 189 different names. According to the family books in 1900-1990
(\tref{table:book}), this number amounts to $m \approx 1.4
\times 10^5$ [\fref{fig:nm}].
Roughly a half of this group is male and if we follow the
lineage backward in time, the remaining lineage will be roughly proportional
to the total population so that $m(t)/M(t)=const$. As the population
decreases, the average number of names, $n$, is hence just given by $n(m)$
provided that this to good approximation is a unique function independent of
time. In \fref{fig:history}(a), the crosses show the prediction from this
assumption, using the information on the total population given in
\fref{fig:estimate} and the estimated $N(M)$ given in \fref{fig:nm}.
As seen here, both the sharp decrease from 2000 to 1900 and the slow decrease
between 1900-1400 are correctly reproduced. In \fref{fig:history}(b), we
do the inverse, using the same assumption:
the data for the 189 families given in \fref{fig:history}(a) is used as
an input to estimate the total population within the period 500-2000.
The full curve in \fref{fig:history}(b) gives the estimated population by
this method. Comparison with the historical estimates from
\cite{history} shows that the agreement is everywhere within a factor
of 2 [see \fref{fig:history}(b)]. The estimate suggests that the
population at about 500 AD was around $4 \times 10^4$ and exceeded $10^7$
around 1400 AD.
It should be noted that our estimate of the total population refers to all
persons integrated in the society having a Korean family name. This might,
of course, in the old age only be part of the actual population controlled
by the society.
\Fref{fig:history}(b) also shows the expected number of family names
based on the prediction for the total population possessing Korean family
names. According to this estimate
there was already around 150 family names around year 500 AD.  Note that the
expected number also reflects the fact that family names can both be added
and subtracted from the population in the cause of history (some families
simply goes extinct). Thus the rate of increase for the total population is
expected to be slower than for the 189 families which only include
families which have survived until year 2000. This observation is also
consistent with our prediction: there are introduced 155 names
during the period 500-2000 according to our data set plotted in
\fref{fig:history}(a), and our predicted increase of the total number
during the same period also happens to be 155.

\begin{figure}
\begin{center}
\includegraphics[width=0.45\textwidth]{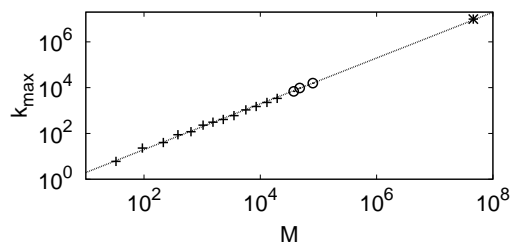}
\end{center}
\caption{The number of persons with the most frequent family name in each
family book. The circles on the upper right side show the three most recent
data sets of \tref{table:book}, from which the slope $a$ of the line has
been determined by linear fitting, $k_{\rm max} = a M$. Note that these
three points are not in time-order since the middle point is the latest. The
crosses are the remaining thirteen time windows for which time and size is in
the same order. The asterisk is the number of Kims for the whole population
according to the census in year 2000. The proportionality with the size of
the group and the number of Kims are borne out to very high degree
irrespective of time.}
\label{fig:kmax}
\end{figure}

\begin{figure}
\begin{center}
\includegraphics[width=0.45\textwidth]{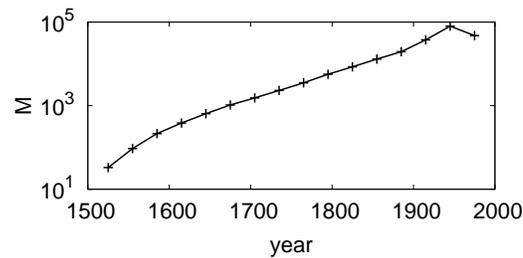}
\end{center}
\caption{$M$ as a function of time in \tref{table:book}.}
\label{fig:famsize}
\end{figure}

A second feature of the RGF model is that the largest group is always
proportional to the total size of the data set~\cite{zipf}. In the context of
Korean family names, this implies that the proportion of persons named `Kim'
in a randomly picked group of Koreans should be constant, irrespectively of
historical time or size of the group. In order to test this, we again start
from the three latest historical windows in \tref{table:book}, i.e.,
1900-1930,1930-1960, and 1960-1990. Note that period 1930-1960 is the
largest group, so that $M$ actually decreases with time during some period
within 1930-1990. These three data points are plotted in \fref{fig:kmax}
(open circles) and the straight line in the figure is obtained by the least
square fitting to these three data points. This linear prediction, based on
the data from 1900-1990, is then directly compared to the data for the
thirteen time windows from 1510 to 1900. The agreement is striking. In
addition, the numbers of Kims from the census of year 2000 is given by the
asterisk. Thus the figure spans group sizes from $M=33$ to $M=4.6\times 10^7$
and history from year 1510 to 2000. From this perspective, the
proportionality is borne out to an amazing degree. Is this a surprising
result? All it is saying is that the total number of persons named Kim
grows and decreases at precisely the same rate as the total population. If
the number of persons belonging to \emph{any} family grows and decreases precisely
as the total population then the result will follow. However, this is not
the case. The ten families in \tref{table:book} show a rapid
increase of inflow of women by marriage from 1510 to 1900. The number of
in-flowing women is an approximate measure of the total number of persons who
carry one of the ten names. \Fref{fig:famsize} shows this rapid
growth: a factor of about 60 to be compared to a factor of about 2 for the
total population during the same period. Thus individual name groups, in
general, grow and decrease in a very different way compared to the total
population. Nor do growth models in general predict that the largest group
grows linearly with the size. For example, the Simon model predicts that the
largest group grows like $k_{\rm max}\propto M^{1-\alpha}$ where $0<\alpha<1$
is the probability for a new name appearing during a
time step~\cite{simon,tr}. This means that only in the trivial case when the
whole population is named Kim (which corresponds to $\alpha=0$) is there a
direct proportionality. Thus, in spite of the simplicity of the result, it
appears to be nontrivial. Nevertheless, it is consistent with the
prediction of the RGF model. We also note that combining our population
estimate of persons with Korean family names with the size proportionality
of Kim suggests that at around 500 AD there were already about 10000 Koreans
named Kim.

\begin{figure}
\begin{center}
\includegraphics[width=0.45\textwidth]{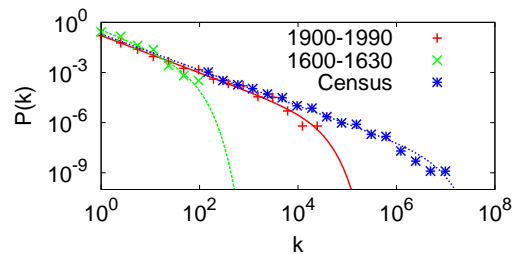}
\end{center}
\caption{Comparison between the actual family books (points)
and RGF predictions (lines).}
\label{fig:pk}
\end{figure}

\begin{figure}
\begin{center}
\includegraphics[width=0.45\textwidth]{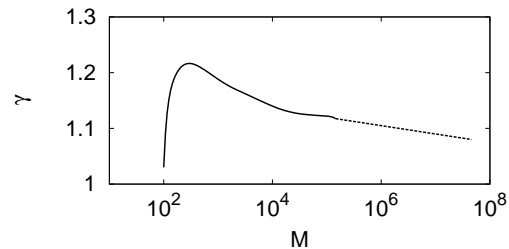}
\end{center}
\caption{Power-law exponent as a function of $M$. The solid line is obtained
by analyzing the family books from 1900-1990, and the dotted line connects to
the census data in 2000.}
\label{fig:gamma}
\end{figure}

So far, we have only discussed two features of the RGF model: the uniqueness
and time independence of the function $N(M)$ and the proportionality
between the size of the largest group for any random part of a population
with time-independent proportionality. However, the RGF model also gives a
precise prediction for the actual group-size distribution $P_M(k)$ in the
form of \eref{eq:pk}. In order to test this prediction, we again start
from the data in \tref{table:book} and the period 1900-1990 which
contains $M_{\rm tot}=165020$ women each having one family name out of
$N_{\rm tot}=194$ and where $k_{\rm max}=32316$ are named Kim. These three
numbers $M_{\rm tot}$, $N_{\rm tot}$
and $k_{\rm max}$ uniquely determines $P_M(k)$ within the RGF model, as
explained in \sref{sec:rgf}~\cite{zipf}. The middle full curve in
\fref{fig:pk} gives the predicted size distribution and the pluses denote
the actual (binned) data points. The agreement between the RGF prediction
and the data is very good, in particular in view of the fact that the
prediction is based solely on the three numbers $M_{\rm tot}$, $N_{\rm tot}$ and
$k_{\rm max}$. The prediction for the exponent $\gamma$ in \eref{eq:pk} is
$\gamma=1.12$. As explained in \sref{sec:rgf}, the RGF model allows you to
predict how the $P_M(k)$ for either a smaller, $m<M$ or larger $m>M$.
\Fref{fig:gamma} displays the predicted change of $\gamma$ when
starting from the data given for the period between 1900-1990. The solid
curve gives the change when $m$ is decreasing and the dotted line as it is
increasing. The fact that $\gamma$ changes with the size of the data set is
a fundamental feature of the RGF model and distinguishes it from the usual
growth models which in general give scale-invariant and hence
size-independent $\gamma$~\cite{zipf}. The left full curve in
\fref{fig:pk} is the prediction for the 1600-1630 data only using the
data for $1900-1990$ and the number $m=384$, which is the number of women
getting married into the ten families during the period 1600-1630.  The
actual name-frequency distribution for these women is given by the crosses
and the agreement is again quite good. In particular, note that the data is
indeed consistent with the slightly steeper slope for smaller $k$ caused by
a slightly larger $\gamma=1.22$ (compare \fref{fig:gamma}). The rightmost
curve in \fref{fig:pk}, in the same way, gives the prediction based on
only using the data for the married women in $1900-1990$ and the total
population size in year 2000 given by  $m= 4.6 \times 10^7$. The census data
from year 2000 is also plotted and the agreement between prediction
and the data is again very good. This time the $\gamma=1.07$ is even closer
to one. In short, the RGF model gives very good predictions for the
distribution of actual Korean name-group sizes both backward and forward in
time.

\section{Concluding Remarks}
\label{sec:conclude}

Our analysis suggests that the family-name distribution within the Korean
population shares characteristic features with the word-frequency
distribution for an author: both are to good approximation described by a
``meta-book'' distribution $N(M)$ and both are well described by the
RGF model~\cite{meta,zipf}. The ``meta-book'' distribution for an author gives
a unique relation between a text of length $M$ written by an author and the
number of distinct words used, $N$. In the Korean case, this corresponds to
the number of distinct family names $N$ you typically find in a group of $M$
Koreans. In the word-frequency case, this leads to the conclusion that the
most common word used by an author in an English text, \emph{the}, is
proportional to the total size of the text $M$. The corresponding conclusion
for the Korean family names is that the most common name, \emph{Kim}, in a
group of $M$ Koreans should on the average always be proportional to a the
size $M$. This prediction was checked with data ranging from 1510 AD to 2000
AD and group sizes in the interval $[33, 4.6\times 10^7]$ and was found to be
obeyed with high precision. It was argued that this is a nontrivial result
for two reasons: first, it was shown that the rise and fall of individual
families in general have no simple relation to the rise and fall of the total
population; the only obvious relation is that the members of all the
families collectively varies as the total population. Second, usual growth
models, like the Simon model, predicts that the size of the largest family
grows slower than the population.

The fact that the name distribution to good approximation appears to follow
a unique $N(M)$, made it possible to estimate the size of the Korean culture
down to year 500 AD, by using the statistics for the years in which 189 Korean
family names were introduced. This estimate suggested that the total
population was around $5 \times 10^4$ persons with Korean family names of
which about $10^4$ carried the family name Kim. The total number of family
names at year 500 AD was predicted to be around 150. We think that these are
fascinating conclusions, although we cannot judge the historical realism and
implication of the ten thousand Kims.

Finally, we demonstrated the actual frequency distributions Korean names
follow the RGF distribution with a size-dependent power-law exponent
$\gamma$~\cite{zipf}.

What does these results imply for the Korean culture? We speculate that the
answer is stability. It seems that some core of the Korean culture has
remained intact over at least 1500 years and as both the population and
occupied area expanded, it basically swallowed other cultural influences
without compromising its core.
An interesting question is if this type of analysis could also be applied to
other cultures. This, however, remains for the future.

\ack
S.K.B and P.M. acknowledge the support from the Swedish Research Council
with Grant No. 621-2008-4449.
B.J.K.  was supported by Basic Science Research Program through the
National Research Foundation of Korea (NRF) funded by the Ministry of
Education, Science and Technology (2010-0008758).

\section*{References}

\end{document}